# Numerical Simulations of an $SU(2)_L \otimes SU(2)_R$-symmetric Higgs-Yukawa Model on the QUADRICS Q16

M. Plagge[a] [*] and D. Talkenberger[b] [†]

[a]Fakultät für Physik, Universität Bielefeld, Universitätsstr. 25, 33615 Bielefeld, Germany

[b]Institut für Theoretische Physik I, Universität Münster, Wilhelm-Klemm-Str. 9, 48149 Münster, Germany

We report on our work on the $SU(2)_L \otimes SU(2)_R$ symmetric Higgs Yukawa Model with mirror fermion action. Our model describes a fermion Higgs system in the limit of vanishing gauge coupling. Setting the bare Yukawa coupling of the mirror fermions $G_\chi$ to zero, we want to determine the triviality bounds on the renormalized Yukawa coupling of the fermions $G_{R\psi}$ and the scalar self-coupling $g_R$ on $8^3 \times 16$ and $16^3 \times 32$ lattices.

## 1. Introduction

Fermion Higgs models with left right symmetric action have been intensively studied over the last years (see [1] and references therein). The results obtained so far are in good agreement with one loop perturbation theory but there are still strong finite size effects, especially at large values of the Yukawa coupling.

With increasing computer power computations on larger lattice sizes become feasible and we can get a better control over the finite size effects (beside the use of improved actions). The use of massivly parallel computers boosts this development. The QUADRICS Q16 is one of these which has a very appealing price/performance ratio.

In the following we report on our progress in implementing the model which has been studied in [1] on the Q16. We present first timing results of the program with dynamical fermions and details of the implementation.

## 2. The Model

The lattice action of the model consists of a pure scalar part and a mixed fermion scalar part $S = S_\varphi + S_\Psi$. The scalar field $\varphi$ is represented as a $2 \otimes 2$ matrix $\varphi = \phi_4 + i\sigma_k \phi_k$ $k = 1, 2, 3$, where $\sigma_k$ are the Pauli matrices and $\phi_\sigma$ are real variables. $\Psi$ is the fermion pair of fermion $\psi$ and mirror fermion $\chi$ doublets, $\Psi = (\psi, \chi)$. Using conventional normalizations we have

$$S_\varphi = \sum_x \left\{ \frac{1}{2} Tr(\varphi_x^+ \varphi_x) + \lambda \left[ \frac{1}{2} Tr(\varphi_x^+ \varphi_x) - 1 \right]^2 \right.$$
$$\left. -\kappa \sum_{\mu=0}^{4} Tr(\varphi_{x+\hat\mu}^+ \varphi_x) \right\}$$

$$S_\Psi = \sum_x \left\{ \mu_{\psi\chi} \left[ \overline\chi_x \psi_x + \overline\psi_x \chi_x \right] \right.$$
$$-K \sum_{\mu=-4}^{4} \left[ \overline\psi_{x+\hat\mu} \gamma_\mu \psi_x + \overline\chi_{x+\hat\mu} \gamma_\mu \chi_x \right.$$
$$\left. +r \left( \overline\chi_{x+\hat\mu}\psi_x - \overline\chi_x\psi_x + \overline\psi_{x+\hat\mu}\chi_x - \overline\psi_x\chi_x \right) \right]$$
$$+ G_\psi \left[ \overline\psi_{Rx}\varphi_x^+ \psi_{Lx} + \overline\psi_{Lx}\varphi_x\psi_{Rx} \right]$$
$$\left. + G_\chi \left[ \overline\chi_{Rx}\varphi_x\chi_{Lx} + \overline\chi_{Lx}\varphi_x^+\chi_{Rx} \right] \right\}$$
$$= \sum_{x,y} \overline\Psi_y Q(\varphi)_{yx} \Psi_x$$

For the Hybrid Monte Carlo the number of flavours has to be doubled so that the last sum becomes a sum over the lattice points and the two flavours.

The choice of the seven parameters follows [1]. The Wilson parameter $r$ is set to 1. In order to study the triviality bound, we set the quartic coupling $\lambda$ to infinity. We choose $\mu_{\psi\chi} = 0.0$ and $G_\chi = 0.0$ to ensure the decoupling of the mirror fermions in the continuum limit. The fermionic hopping parameter $K$ is fixed to $K_{cr} = 0.125$ [1]. This leaves the fermionic Yukawa coupling $G_\psi$ and the hopping parameter $\kappa$ as physical input parameters. We use $\kappa$ to tune the scalar mass and select the phase.

[*]Supported by Deutsche Forschungsgemeinschaft under grant PE 340/3-2
[†]Supported by Deutsche Forschungsgemeinschaft under grant Mu 757/4-3

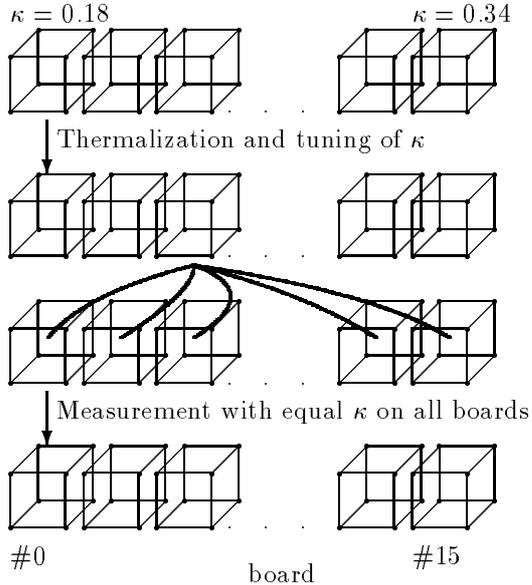

Figure 1. Example run on the Q16

## 3. Implementation on the QUADRICS

To simulate the model we use the hybrid Monte Carlo algorithm with conjugate gradient for the matrix inversion. For information on the parameter dependence we refer to [1],[4] and [5].

The topology of the Q16 is $2 \times 2 \times 32$ which is not well suited for a lattice of size $8^3 \times 16$. But it is possible to run the Q16 as 16 independend Q1 boards with $2 \times 2 \times 2$ topology and periodic boundary conditions closing on each board. Therefore we decided to implement the program in a sort of two step parallelism. First we did a geometric parallelization on the Q1 boards in dividing the lattice in 8 parts and putting each one on of the 8 CPUs. The second step is simply to run a different configuration on each of the 16 Q1 boards. The drawback of this approach seems to be that one has to equilibrate each configuration and so waste a lot of CPU time. Fortunately, this is not true in our case. Before we can start our measurements, we have to tune the scalar hopping parameter $\kappa$ in order to achieve a scalar mass of $0.6 - 0.8$ in lattice units and to choose the desired phase (here the broken, FM, phase, see [1]). Fig. 1 shows what we do. We start on each Q1 with a different value of $\kappa$, thermalize these configurations and then determine the scalar mass on the 16 boards. What we get is shown in Fig. 2. From this picture we determine the hopping parameter value for the measurements. We pick out the configuration which lies closest to this values, duplicate it 16 times and restart the program now with 16 equal configurations and $\kappa$ values. If the new $\kappa$ value lies in between two old $\kappa$ values we do again a few thermalization sweeps, the number of which is of the order 10 smaller than on the original start configurations. At the end we get 16 independend results from which determine the quantities of interest by the bootstrap method. Within each configuration the errors are estimated by a jackknife approach.

The larger $16^3 \times 32$ lattice will be run on the DFG QH2 which will allow a more flexible partitioning, so that we can use the same method but with less configurations.

A problem which has to be faced on the QUADRICS is its single precision floating point hardware. There are three points in the program where global sums have to be calculated and where the low precision can cause problems. In the metropolis update at the end of each trajectory small errors in the summation produce rather big effects on the update probability due to the exponential function involved. Errors in the scalar products, which have to be evaluated during the conjugate gradient iterations, influence the convergence in an unpredictable manner. And, last not least, in the measurement routine, where we expect the errors to play a less important role. We have decided to use in all these cases a summation procedure where the terms are added tree like, so that at every step terms of equal size are added.

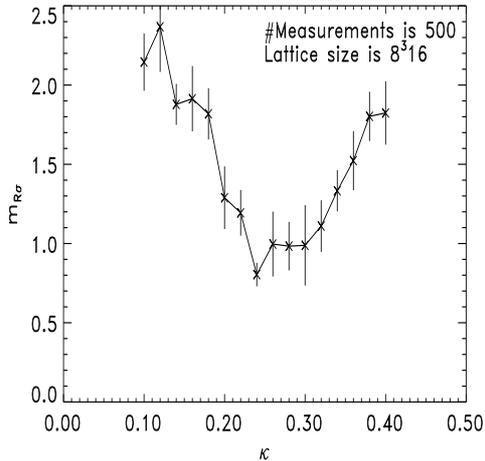

Figure 2. $\kappa$ tuning at $G_\psi = 0.3$

For the random numbers we use the internal rand16 library routine, so we have for each of the 128 CPUs an independend generator.

A QUADRICS feature which caused some problems is the limitation on the size of the executables. This limit lies at $\approx$ 2.6MB, a size which can be reached very quickly by using compile time for loops even if the length of these loops is small (in our program typically 2-4).

### 4. Timing results

The most time consuming part in the simulation is the conjugate gradient. It utilizes almost 95% of the whole CPU time. We have compared the times for a single matrix inversion with times from a CRAY YMP. For a fixed number of iterations we are about 10% faster on the Q1. If we use the residual as the stopping criteria, which is of course more realistic, we are about 10% slower, which results from an increase in the number of iterations. We think this increase is due to summation errors in the scalar product, but we will have to clarify this point a bit more, before going to larger lattices. Compared to the peak performance we achieved $\approx$ 45% in the conjugate gradient. These values are in good coincidence with the results obtained with the performance analyzer.

Lattices of size $4^3 \times 8$ can be run either on a single CPU or, geometrically parallelized, on a eight CPU Q1 board. We have measured a speed up factor of 6 for one hybrid trajectory.

Comparing the times for different lattice sizes on the Q1, an increase in the performance, measured in MFLOPS, with increasing lattice size can be seen. This is due to a better filling of the pipeline.

To sum up, we can say that the Q16 gives for our problem an increase in computer power of a factor 15 compared to a CRAY YMP. This allows us to achieve high statistics in a reasonable time on lattice sizes up to $16^3 \times 32$. The main difficulties are the limitation on the code size and the single precision arithmetic.